\documentclass[12pt,preprint]{article}
\usepackage{epsf}
\newcommand{\eqb}{\begin{equation}}
\newcommand{\eqe}{\end{equation}}
\newcommand{\dmb}{\begin{displaymath}}
\newcommand{\dme}{\end{displaymath}}

\newcommand{\ep}{\varepsilon}
\newcommand{\eab}{\begin{eqnarray}}
\newcommand{\eae}{\end{eqnarray}}
\newcommand{\ra}{\right\rangle}
\newcommand{\la}{\left\langle}
\newcommand{\e}{\mbox{e}}
\newcommand{\be}{\begin{equation}}
\newcommand{\ee}{\end{equation}}

\begin{document}
\begin{titlepage}
\begin{flushright}
MPI-PhT 2001-33 \\
\end{flushright}
\vspace{0.6cm}

\begin{center}
\Large{{\bf Towards OPE based local quark-hadron duality: Light-quark channels}}

\vspace{1cm}

Ralf Hofmann

\end{center}
\vspace{0.3cm}

\begin{center}
{\em Max-Planck-Institut f\"ur Physik\\ 
Werner-Heisenberg-Institut\\ 
F\"ohringer Ring 6, 80805 M\"unchen\\ 
Germany}
\end{center}
\vspace{0.5cm}

\begin{abstract}

Various light-quark channel current-current correlators are subjected to the 
concept of a non-perturbative component of coarse graining 
in operator product expansions introduced in a parallel work. 
This procedure allows for low-energy structure of the OPE-derived spectral function. 
With naive vacuum saturation for 4 quark operators and using lattice data for the gauge invariant scalar 
quark correlator the results are far off the experimentally measured behavior. 
However, using the correlation length of the gauge invariant vector 
quark correlator, which is about 10 times smaller than the scalar one, 
the qualitative results are rather realistic. Namely, 
the input of information on the mass of the lowest 
resonance in one channel yields the corresponding masses 
within acceptable errors in other channels. Still, the shapes of 
the calculated spectral functions are 
considerably deformed as compared to experiment. 
This may be a consequence of vacuum saturation 
and the truncation at a mass dimension which is 
below the critical dimension from which on the asymptotic 
expansion does not approximate anymore. To improve on this 
high-resolution lattice information on gauge invariant 
$n>2$ point correlators would be needed. 
Motivated by the small effective 
correlation length in the 4 quark contributions 
the relevance of the approach for heavy quark physics, in particular 
in the calculation of nonleptonic, 
inclusive $\Delta \Gamma$, is discussed.

\end{abstract} 

\end{titlepage}

\section{Introduction}

The basic assumption made in QCD sum rules (QSR) in particular \cite{SVZ} and in 
applications of the operator product expansion (OPE) 
to hadronic physics in general \cite{Bigi} is 
quark-hadron duality, namely the possibility to express the low-energy dispersive 
part of a correlator of QCD currents in terms of hadronic cross sections. 
For external momentum $q$ with $q^2=-Q^{2}<0$ correlators are believed to 
be (asymptotically) approximated by the OPE which is a power series in 
$Q^{-2}$. A first estimate for the critical dimension 
$d_c\sim 12$ from which on the entire coefficients, 
being products of Wilson coefficients and averages over local, gauge invariant operators,  
are dominated by short distance effects was performed in ref.\,\cite{SVZ} for pure gluodynamics. 
Thereby, the vacuum was described by a dilute instanton gas. Powers with $d>d_c$ 
have no potential to reflect the hadronic properties of the 
respective channel which relate to the operator 
averages via dispersion relations in QSR (global duality) 
and via the optical theorem and analytical continuation in applications such as calculations 
of nonleptonic decay widths of heavy mesons (local duality).

Apart from some puzzles\footnote{large channel-to-channel variation of 
spectral continuum thresholds, see ref.\,\cite{CNZ}, large error in vacuum averages of local operators, 
unclarified status of such approximations as vacuum saturation in 
vacuum averages over 4-quark operators, very large non-perturbative effects 
in the scalar and pseudoscalar channels} 
the tremendous success of the sum rule 
method seems to support the global version of quark-hadron duality. 

The observation that an analytical continuation of the OPE to time-like, 
external momenta does not yield the phenomenological resonance 
structure in the imaginary part has lead to the 
notion of local duality {\sl violation} \cite{Bigi,Shifman0,Shifman1}. 
The up-to-date claim is that the asymptotic nature of the 
expansion does not allow for additive, 
exponential-like terms to appear. These 
terms are to be responsible for 
a ``wiggling'' of the low-energy part of the spectral function \cite{Bigi}. 
This, however, seems to contradict the QSR philosophy \cite{SVZ}: If only the first few terms of the OPE 
(up to the critical dimension $d_c$) are needed to 
describe the lowest resonances in a given channel via a dispersion relation then 
operators of dimension $d\ge d_c$ should also be forgotten 
in an OPE-based construction of the spectral function.

Therefore, we propose an alternative in \cite{PRL}. The observation is that 
the non-perturbative behavior of the QCD vacuum, when probed with low euclidean momenta, 
is not only characterized by nonvanishing vacuum expectation 
values (VEV's) of local, gauge invariant operators but also 
by {\sl finite} correlation lengths of the 
corresponding gauge invariant correlators \cite{Dosch,DiGiacomo}. 
As a consequence the fundamental field operators must loose their relevance 
with decreasing resolution (see next section). From the knowledge of gauge invariant 
correlation functions of fundamental operators at high resolution $Q_0$ evolution 
equations for the VEV's $A(Q)$ of the relevant local, effective 
operators can be derived for $Q<Q_0$. For $d=4$, where 2-point functions are needed, 
we have $A(Q)\sim \exp[4/5(\lambda Q)^{-1}]$ with 
$\lambda$ being the correlation length. Compared with the conventional power correction 
the effect becomes noticable if $Q\sim \lambda^{-1}$. For gluon and quark condensates 
the lattice implies $A(Q_0)$'s at 
$Q_0\sim 2$ GeV which are compatible with the 
phenomenological values obtained from conventional QSR's at $\mu\sim$ 1 GeV. 
Since the inverse correlation lengths are below the mass of the 
$\rho$-resonance (assuming naive vacuum saturation at dimension 6) 
light-quark channel QCS sum rules are hardly 
touched by the exponential decrease of the operator VEV's \cite{PRL}. 
However, calculating the spectral function from the truncated 
OPE these large correlation lengths do not reflect the resonance
physics in the respective channels. A much smaller, 
effective correlation length is needed at dimension 6 to 
yield more realistic spectra. 
To investigate the properties of OPE-based spectral functions in various light-quark 
channels is the main purpose of this paper.

The presentation is set up as follows: In the next section we briefly review the idea of non-perturbative 
coarse graining of local operator VEV's as it is developed in \cite{PRL}. Thereby, 
the focus is on correlation functions which are represented by so-called connected diagrams.  
Section 3 investigates the vacuum saturation hypothesis for 4-quark 
operators \cite{SVZ} in the light of non-perturbative coarse graining. The low-energy parts of the 
spectral functions in the $\rho$, $a_1$, $\pi$, and $\phi$ channels are 
calculated in section 4 by analytical continuation of the 
OPE to time-like external momenta. Implications 
for the calculation of the difference of nonleptonic inclusive 
decay widths in neutral $B$-meson systems are discussed in section 5. 
The last section summarizes the results.

\section{Euclidean exponentials and Minkowskian oscillations}

In this section we briefly review the work of \cite{PRL} on non-perturbative coarse graining for 
VEV's of local, gauge invariant operators. 

At a large euclidean momentum $Q\equiv\sqrt{Q^2}$, where we expect the 
description of the dynamics in terms of the continuum action and local operators made of {\sl fundamental} fields 
to be sufficiently accurate, we start by expanding the current-current correlator into 
a conventional OPE. The evolution to lower momenta is obtained by running the Wilson coefficients 
{\sl perturbatively} via the running coupling $\alpha_s(Q^2)$ and the 
anomalous operator dimensions \cite{SVZ}. According to \cite{PRL} 
the {\sl non-perturbative} 
running of an operator average is governed by the non-perturbative part of the corresponding 
gauge invariant correlator in euclidean position space. This correspondence can be expressed as
\eqb
\label{nl}
\la F(0)\ra_Q^{np} \equiv \la F_1(0)\cdots F_n(0)\ra_Q^{np} \to 1/{\cal N}\sum \la F_1(0)\cdots F_n(x_n)\ra_Q^{np}\ . 
\eqe
In eq.\,(\ref{nl}) parallel transporters 
\eqb
S(0,x)\equiv{\cal P}\exp\left[ig\int_0^{x}dz_\mu\, A_\mu\right]
\eqe
are appropriately contained in the non-local 
expression to define gauge invariant correlations. 
The sum runs over all relevant\footnote{This will be specified later.}, piecewise straight \cite{Dosch} 
trajectories of parallel transport, and the symbol ${\cal P}$ demands path ordering. 
With the normalization factor $1/{\cal N}$, which depends on $n$ and the numbers of fields transforming 
under the fundamental and adjoint representation, the correlation function reduces 
to the ``condensate'' in the limit $x_1,\cdots,x_n\to 0$. Making the convention that 
an arrow pointing from $x_i$ towards $x_j$ stands for 
the parallel transport $S(x_i,x_j)$, we have a way to depict correlators. Note that points with fields 
transforming under the fundamental (adjoint) representation are 
connected to one (two) lines of parallel transport. There are disconnected and connected diagrams. 
In this section we only consider the latter. 

To coarse grain from resolution $Q$ to resolution $Q-dQ$ we 
average the non-perturbative part of 
the correlator corresponding to a connected diagrams over a (euclidean) 
ball of radius $dR_Q$ with
\eqb
dR_Q=\frac{1}{Q-dQ}-\frac{1}{Q}=\frac{1}{Q}\left(\frac{1}{1-\frac{dQ}{Q}}-1\right)\sim \frac{dQ}{Q^2}\ .
\eqe
Using the short-hand notation of eq.\,(\ref{nl}), this is written as
\eqb
\label{gc}
\la F(0)\ra_{Q-dQ}^{np}=\frac{1}{(V(dR_Q))^{(n-1)}}\int_{|x_1|,\cdots,|x_n|
\le dR_Q}d^4x_1\cdots d^4x_n \la F_1(0)\cdots F_n(x_n)\ra_Q^{np}\ ,
\eqe
where
\eqb
V(dR_Q)=\frac{1}{2}\pi^2 (dR_Q)^4=\frac{1}{2}\pi^2\,\left(\frac{dQ}{Q^2}\right)^4\ .
\eqe
In Fig.\,1 there are 3 examples for connected diagrams.
\begin{figure}
\vspace{5.3cm}
\includegraphics{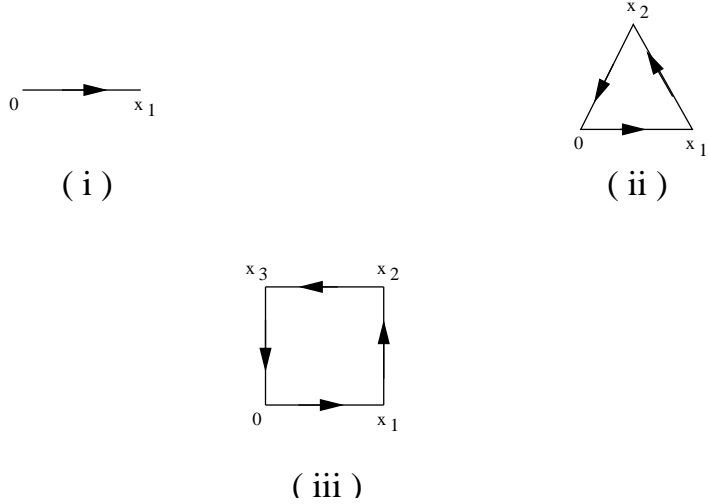}
\caption{Diagrammatic representation of the gauge invariant correlators corresponding to 
the local operators $(i)$ 
$\bar{q}(0)q(0)$, $(ii)$ $\mbox{tr} F_{\mu\nu}(0) F_{\nu\kappa}(0) F_{\kappa\mu}(0)$, 
and $(iii)$ $\mbox{tr} F_{\mu\nu}(0) F_{\nu\kappa}(0) F_{\kappa\lambda}(0) F_{\lambda\mu}(0)$.} 
\label{} 
\end{figure}

Let us now focus on 2-point functions as they are relevant for dimension 3 and 4 
quark and gluon operators, respectively. Only the gauge invariant 
bilocal quark correlator \cite{DiGiacomo}
\eqb
\label{qco}
\la\mbox{tr}\,\bar{q}(x)\,S(x,0)\,q(0)\ra
\eqe
and the gluonic field strength correlator \cite{Dosch,DiGiacomo}
\eqb
\label{gco}
\la\mbox{tr}\,F_{\mu\nu}(x)\,S(x,0)\,F_{\kappa\lambda}\,S^\dagger(x,0)\ra
\eqe
have been measured on the lattice. The results 
imply that there exists an additive decomposition into a perturbative, power-like in $|x|$, 
and a non-perturbative, exponential in $|x|$ piece \cite{DiGiacomo}. We are interested in the 
non-perturbative part, which in short-hand reads
\eqb 
\label{ex}
\la F_1(0)F_2(x)\ra_Q^{np}=A(Q)\,\exp(-|x|/\lambda)\ .
\eqe
As explained in \cite{PRL} $\lambda$ is expected to have direct 
phenomological meaning, and so 
it can not depend on the resolution $Q$. Hence, coarse graining may 
only affect the pre-exponential factor $A(Q)$. 
An evolution equation for $A(Q)$ can be derived if we combine eqs.\,(\ref{gc}) and (\ref{ex}):
\eqb
\label{gc2}
\left(V(dQ/Q^2)\right)^{-1}\,\int_{|x|\le \frac{dQ}{Q^2}}d^4x\, A(Q)\e^{-|x|/\lambda}=
A(Q-dQ)\e^{-0/\lambda}=A(Q-dQ)\ .
\eqe
Note that for $\lambda\to\infty$ $A$ is invariant under coarse graining. Therefore, 
the conventional treatment of non-perturbative corrections 
in the framework of the OPE \cite{SVZ} is recovered, and condensates do not 
depend on the resolution in this limit. In particular, they could be 
determined at a large resolution, where 
fundamental fields make sense. In the real world, however, 
correlation lengths {\sl are} finite \cite{DiGiacomo}. 
We will see later how this is reflected in the existence of hadronic resonances.

Expanding the l.h.s. and r.h.s. of eq.\,(\ref{gc2}) in $\frac{dQ}{Q^2}$ and comparing coefficients 
of the linear terms, we obtain the following equation:
\eqb
\label{evo}
\frac{d}{dQ}\,A(Q)=\frac{4}{5\lambda}\,Q^{-2}A(Q)\ ,
\eqe
with solution
\eqb
\label{run}
A(Q)=A(Q_0)\exp\left[-\frac{4}{5\lambda}\left(\frac{1}{Q}-\frac{1}{Q_0}\right)\right]\ .
\eqe
At dimension 4 there are no anomalous operator dimensions and being 
content with a determination of Wilson coefficients at the lowest possible order in $\alpha_s$ 
the generic form of a non-perturbative correction is
\eqb
\label{pc}
\frac{A_4(Q_0)}{(Q^2)^2}\,\exp\left[-\frac{4}{5\lambda_4}\left(\frac{1}{Q}-\frac{1}{Q_0}\right)\right]\ .
\eqe
Eq.\,(\ref{run}) implies that compared to conventional dimension 4 power corrections there is a noticable suppression 
if $Q$ is of the order of $\lambda^{-1}$ or less.  

Lattice measurements with $N_F=4$ staggered fermions of mass $a\cdot m_q=0.01$ and 
lattice resolution $Q_0=a^{-1}\sim 2$ GeV suggest that scalar fermionic and gluonic correlation 
lengths are $\lambda^s_q\sim$ 3.2 GeV$^{-1}$ and $\lambda_g\sim 1.7$ GeV$^{-1}$, 
respectively \cite{DiGiacomo}. For the vector fermionic correlation length 
$\lambda^v_q\sim 1/10\, \lambda^s_q$ was found \cite{DiGiacomo}. 
So for dimension 4 and, assuming naive vacuum saturation, also for dimension 6 this does 
not seem to pose a problem for the sum rule analysis of light quark correlators (at 
$Q_0=a^{-1}\sim 2$ GeV $A_q(Q_0)$ as well as $A_g(Q_0)$ are 
compatible with their QSR determined values at $\mu\sim 1$ GeV \cite{DiGiacomo}). 
However, as we will see later, naive vacuum saturation gives light-quark channel spectra which 
are completely off the experimentally measured behavior. 
Qualitatively more realistic spectral functions can be obtained using the much 
smaller $\lambda^v_q$.

The contribution of dimension 4 to the spectral function $\rho(s)$ 
is (up to a normalization!) 
obtained by analytically continuing to $Q^2=-(s+i\ep)$ or $Q=-i\sqrt{s}$, $(s>0)$, 
and taking the imaginary part. 
A term like the one in eq.\,(\ref{pc}) corresponds to a term 
\eqb
\label{sf}
\frac{A_4(Q_0)}{s^2}\exp\left[\frac{4}{5\lambda_4}\frac{1}{Q_0}\right]
\sin\left[-\frac{4}{5\lambda_4 \sqrt{s}}\right]
\eqe
in $\rho(s)$. The oscillatory behavior manifests itself in a quite different way than 
it was suggested in refs.\,\cite{Shifman1}: $\sin[\sqrt{s}^{-1}]$ 
instead of $\sin[\sqrt{s}]$ or $\sin[s]$. There, oscillations are present everywhere though power suppressed at high $s$. 
Here, oscillations only start if $\sqrt{s}^{-1}$ is larger than the 
correlation length of the corresponding $2$-point function.

\section{Vacuum saturation in the context of gauge invariant correlations}

After the treatment of connected diagrams in the previous section we 
turn to disconnected diagrams in this section. 
A sufficient condition for factorized coarse graining is the factorization of the 
non-perturbative part in the corresponding correlation function. 
It is unlikely that such a factorization occurs for connected diagrams. 

Let us focus on the case of 4 quark operators since generalizations to higher dimensions 
are straightforward. Disconnected diagrams are associated with 4 quark operators 
$\bar{q}(0)\Gamma q(0)\bar{q}(0)\Gamma q(0)$ composed 
of color singlet currents (Fig.\,2). Thereby, $q$ is a single flavor quark field, and 
$\Gamma$ denotes one of ${\bf 1},\gamma_5,\gamma_\mu,\cdots$ or a product of them. 
\begin{figure}
\vspace{4.3cm}
\includegraphics{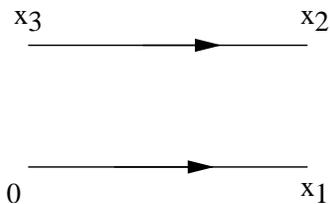}
\caption{Diagrammatic presentation of the gauge invariant correlator corresponding to 
the local operator $\bar{q}(0)\Gamma q(0)\bar{q}(0)\Gamma q(0)$.} 
\label{} 
\end{figure}
Diagrams with 
permutated arguments are irrelevant because of the integrations 
in eq.\,(\ref{gc}) and the fact that we may assume 
translational, $P$, and $T$ invariance \cite{Meg}. In practical applications 
one encounters color octet currents in 
4 quark operators. These structures lead to special 
forms of disconnected diagrams which are due to the fact 
that gauge invariant correlation functions can only be defined 
if 2 of the 4 arguments coincide, hence yielding 3 point functions. 
For example, an operator 
$\bar{q}(0)\Gamma t^a q(0)\bar{q}(0)\Gamma t^a q(0)$ 
demands non-local contributions of the form
\eab
\label{oc}
(i)&&\ \ \la\bar{q}(0) \Gamma S(0,x_1) t^a q(x_1) \bar{q}(x_2) S(x_2,0) \Gamma t^a q(0)\ra\ ;\nonumber\\ 
(ii)&&\ \ \la\bar{q}(x_1) \Gamma t^a S(x_1,0) q(0) \bar{q}(x_1) \Gamma t^a S(x_1,x_2) q(x_2)\ra\ .
\eae
Thereby, the color matrices are normalized as $\mbox{tr}\, t^a t^b=2\delta^{ab}$. 
The corresponding diagrams are depicted in Fig\,3. Lines do not meet at 
the points $0$ (for $(i)$) and $x_1$ (for $(ii)$) because they connect to different fields.       
\begin{figure}
\vspace{5.3cm}
\includegraphics{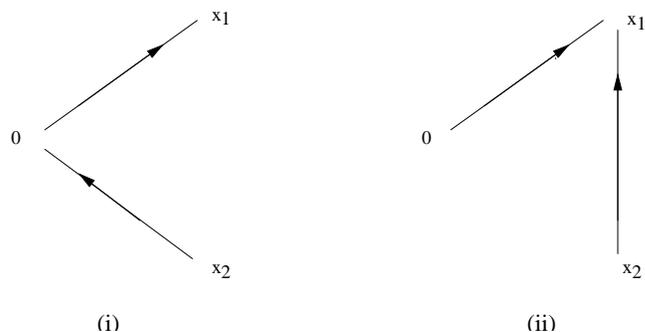}
\caption{Diagrammatic presentation of the contributions of eq.\,(\ref{oc}) to the 
gauge invariant correlator associated with 
the local operator $\bar{q}(0) \Gamma t^a q(0) \bar{q}(0) \Gamma t^a q(0)$.} 
\label{} 
\end{figure}

Since gauge invariant $n>2$ point functions have yet not been measured 
on the lattice we do have to think of approximations 
involving only 2-point functions. The (quite intuitive) hypothesis of vacuum saturation 
can be implemented in two ways:
{\bf 1)} vacuum saturation on the level of {\sl local} operators according to 
the formula of ref.\,\cite{SVZ} with 
separate coarse graining for each factor and {\bf 2)} 
vacuum saturation on the level of the 
correlation function, that is, first delocalization of the 
operator and then vacuum insertion. There is one obvious reason why the 
two prescriptions lead to different results. In case {\bf 1)} we essentially 
square the result for the quark condensate, which implies 
correlations between points of maximal separation $dQ/Q^2$, whereas in 
case {\bf 2)} one factor may contain correlations 
between points of maximal separation $2dQ/Q^2$ (Fig.\,2 or case $(ii)$ in Fig.\,3). 
Another, not-so-obvious reason becomes apparent if we insert the vacuum state into the correlator for 
singlet currents
\eqb
\label{4cor}
\la \bar{q}(0) \Gamma S(0,x_1) q(x_1) \bar{q}(x_2 ) \Gamma S(x_2,x_3) q(x_3) \ra\ .
\eqe
Then information of the 2-point function
\eqb
\label{2cor}
\la \bar{q}(0) \Gamma S(0,x) q(x) \ra
\eqe
is needed. From ref.\cite{Meg} we know that on a 
lattice the correlation length $\lambda_v$ of the vector correlator ($\Gamma=\gamma_\mu$) 
is about 10 times smaller than the correlation length 
$\lambda_s$ of the scalar correlator 
($\Gamma={\bf 1}$)! This result has been obtained for 
a quark mass $m$ with $a\cdot m=0.01$ and $a^{-1}\sim 2$ GeV 
in the $N_F=4$ theory with staggered fermions. 
In the octet case vacuum insertion inbetween the currents 
factorizes the correlator into gauge {\sl variant} 2 point functions. 
The gauge invariant product of them has not yet been measured on the lattice so we can not 
compare the corresponding correlation 
length with the singlet case.

Let us keep all these points in mind in the next section, 
where for simplicity we apply prescription {\bf 1)}. With the necessary 
lattice information available we hope to investigate 
case {\bf 2)} in a separate publication.

\section{Light-quark channels}

After writing down the coarse grained OPE's in the $\rho$, $a_1$, $\pi$, and $\phi$ channels, 
in this section we {\sl calculate} the respective spectral functions by analytical 
continuation from euclidean to time-like momenta. We restrict ourselves 
to light-quark channels because here the two 
needed correlation functions for corrections up to 
dimension 6 (assuming vacuum saturation in the sense of {\bf 1)}) 
have been measured on the lattice. We investigate with two sets of parameters. Set $(A)$ takes the $N_F=4$ results of 
refs.\,\cite{DiGiacomo} for the lowest 
quark mass ($a\cdot m=0.01$) literally, that is
\eab
\label{paraA}
(A):& & \nonumber\\ 
\lambda_q&=&3.1\, \mbox{GeV}^{-1}\ ,\ \ \ \ A_q(a^{-1}=Q_0\sim 2\,\mbox{GeV})=(0.212\, \mbox{GeV})^3\ ;\nonumber\\ 
\lambda_g&=&1.7\, \mbox{GeV}^{-1}\ ,\ \ \ \ A_g(a^{-1}=Q_0\sim 2\,\mbox{GeV})=0.015\, (\mbox{GeV})^4\ .
\eae
Due to the uncertainty in the way dimension 6 contributions are treated and 
motivated by the 10 times smaller vector correlation length $\lambda_v$ we use $\lambda_q=\lambda_v$ in $(B)$
\eab
\label{paraB}
(B):& & \nonumber\\ 
\lambda_q&=&0.3\, \mbox{GeV}^{-1}\ ,\ \ \ \ A_q(a^{-1}=Q_0\sim 2\,\mbox{GeV})=(0.212\, \mbox{GeV})^3\ ;\nonumber\\ 
\lambda_g&=&1.7\, \mbox{GeV}^{-1}\ ,\ \ \ \ A_g(a^{-1}=Q_0\sim 2\,\mbox{GeV})=0.015\, (\mbox{GeV})^4\ .
\eae

\subsection{$\rho$-correlator}

Here, we investigate the correlator of 
currents $j^\rho_\mu=1/2(\bar{u}\gamma_\mu u-\bar{d}\gamma_\mu d)$ 
which at $Q\sim Q_0$ and up to dimension 6 has the following conventional OPE \cite{SVZ}
\eab
\label{rho}
& &i\int d^4 \e^{iqx}\,\la T\{ j^\rho_\mu(x)j^\rho_\nu(0)\}\ra=(q_\mu q_\nu-g_{\mu\nu}q^2)\times\nonumber\\ 
& &\left\{-\frac{1}{8\pi^2}\left(1+\frac{\alpha_s(Q^2)}{\pi}\right)\log\frac{Q^2}{Q_0^2}+
\frac{1}{Q^4}\left[\frac{1}{2}\la m_u\bar{u}u+m_d\bar{d}d\ra_{Q_0}+\frac{1}{24}
\la \frac{\alpha_s}{\pi} F_{\mu\nu}^a F_{\mu\nu}^a\ra_{Q_0}\right]-\right.\nonumber\\ 
& &\left.\frac{\pi\alpha_s(Q^2)}{Q^6}\left[\frac{1}{2}\la (\bar{u}\gamma_\mu\gamma_5 t^a u-\bar{d}\gamma_\mu\gamma_5 t^a d)^2\ra_{Q_0}+
\frac{1}{9}\la 
(\bar{u}\gamma_\mu t^a u+\bar{d}\gamma_\mu t^a d)\sum_{q=u,d,s}\bar{q}\gamma_\mu t^a q\ra_{Q_0}\right]\right\}\ .\nonumber\\ 
\eae
After application of the vacuum saturation hypothesis \cite{SVZ}, going to the SU(2) chiral limit 
$m_u=m_d=0$, and implementing the non-perturbative coarse graining of operator averages\footnote{At dimension 6 there is 
logarithmic running of the Wilson coefficients due to the nonvanishing anomalous dimensions of the corresponding operators. 
Together with the running coupling $\alpha_s(Q^2)$ 
these dependences almost cancel \cite{SVZ}, and hence we will omit them throughout what follows. Operators of dimension 4 
are perturbatively invariant under the renormalization group. We take the 
value $\alpha_s(Q^2_0=(2\,\mbox{GeV})^2)\sim 0.2$. 
This is motivated by ref.\,\cite{Weisz} where a {\sl non-perturbative} 
evolution of $\alpha_s$ has been obtained 
in two flavor lattice QCD using the Schr\"odinger functional scheme. 
With the conversion $\Lambda_{\overline{\mbox{MS}}}=2.382 \Lambda$ and 
taking $\Lambda_{\overline{\mbox{MS}}}=0.4$ GeV, one obtains $\alpha_s(2\,\mbox{GeV})^2)\sim 0.19$. 
We have varied the coupling up to 
$\alpha_s=0.4$ but no drastic changes occur.} 
the restriction $Q\sim Q_0$ can be dropped, and 
the scalar part of the correlator (curly brackets on the r.h.s. of eq.\,(\ref{rho})) 
reduces to
\eab
\label{rhovs}
&&\Pi^\rho(Q^2)=-\frac{1}{8\pi^2}\left(1+\frac{\alpha_s(Q^2)}{\pi}\right)\log\frac{Q^2}{Q_0^2}+
\frac{1}{24 Q^4}\exp[-\frac{4}{5\lambda_g}(\frac{1}{Q}-
\frac{1}{Q_0})]\la \frac{\alpha_s}{\pi} F_{\mu\nu}^a F_{\mu\nu}^a\ra_{Q_0}-\nonumber\\ 
&&\frac{\pi\alpha_s(Q_0^2)}{Q^6}\frac{112}{81}\exp[-\frac{8}{5\lambda_q}(\frac{1}{Q}-\frac{1}{Q_0})]\la\bar{q}q\ra_{Q_0}^2\ .
\eae

\subsection{$a_1$-correlator}

The vacuum averaged OPE for the correlator of the current 
$j^{a_1}_\mu=1/2(\bar{u}\gamma_\mu \gamma_5\, u-\bar{d}\gamma_\mu \gamma_5\, d)$ at $Q\sim Q_0$ and up to dimension 6 
reads in the SU(2) chiral limit \cite{SVZ} 
\eab
\label{a1}
& &i\int d^4 \e^{iqx}\,\la T\{ j^{a_1}_\mu(x)j^{a_1}_\nu(0)\}\ra=(q_\mu q_\nu-g_{\mu\nu}q^2)\times
\left\{-\frac{1}{8\pi^2}\left(1+\frac{\alpha_s(Q^2)}{\pi}\right)\log\frac{Q^2}{Q_0^2}+
\right.\nonumber\\ 
& &\left.\frac{1}{24 Q^4}\la \frac{\alpha_s}{\pi} F_{\mu\nu}^a F_{\mu\nu}^a\ra_{Q_0}-\frac{\pi\alpha_s(Q^2)}{2 Q^6}\la (\bar{u}\gamma_\mu t^a u)^2-
(\bar{u}\gamma_\mu\gamma_5\, t^a u)^2+(u\leftrightarrow d)\ra_{Q_0}\right\}\nonumber\\ \ .
\eae
After vacuum saturation and implementation of non-perturbative operator running the restriction $Q\sim Q_0$ can be forgotten, and 
we have 
\eab
\label{a1vs}
&&\Pi^{a_1}(Q^2)=-\frac{1}{8\pi^2}\left(1+\frac{\alpha_s(Q^2)}{\pi}\right)\log\frac{Q^2}{Q_0^2}+
\frac{1}{24 Q^4}\exp[-\frac{4}{5\lambda_g}(\frac{1}{Q}-
\frac{1}{Q_0})]\la \frac{\alpha_s}{\pi} F_{\mu\nu}^a F_{\mu\nu}^a\ra_{Q_0}+\nonumber\\ 
&&\frac{\pi\alpha_s(Q_0^2)}{Q^6}\frac{176}{81}\exp[-\frac{8}{5\lambda_q}(\frac{1}{Q}-\frac{1}{Q_0})]\la\bar{q}q\ra_{Q_0}^2\ .
\eae

\subsection{$\pi$-correlator}

The pion current is given as $j^{\pi}_\mu=1/2\,i(\bar{u}\gamma_5\, u-\bar{d}\gamma_5\, d)$, and in the SU(2) 
chiral limit its 
correlator can be expanded as \cite{SVZ}
\eab
\label{pi}
&&i\int d^4 \e^{iqx}\,\la T\{ j^{\pi}(x)j^{\pi}(0)\}\ra=-3Q^2\left\{
-\frac{1}{16\pi^2}\left(1+\frac{\alpha_s(Q^2)}{\pi}\right)\log\frac{Q^2}{Q_0^2}-
\frac{1}{48 Q^4}\la \frac{\alpha_s}{\pi} F_{\mu\nu}^a F_{\mu\nu}^a\ra_{Q_0}-
\right.\nonumber\\ 
&&\left.\frac{\pi\alpha_s(Q^2)}{Q^6}\left[\frac{1}{12}\la 
(\bar{u}\sigma_{\mu\nu}\gamma_5 t^a u-
\bar{d}\sigma_{\mu\nu}\gamma_5\, t^a d)^2\ra_{Q_0}+\frac{1}{18}\la (\bar{u}\gamma_{\mu}t^a u+
\bar{d}\gamma_\mu t^a d)\sum_{q=u,d,s}\bar{q}\gamma_\mu t^a q\ra_{Q_0}\right]\right\}\nonumber\\ \ .
\eae
After vacuum saturation and with non-perturbative operator running (no restriction $Q\sim Q_0$ anymore)
the piece in curly brackets becomes
\eab
\label{pivs}
&&\Pi^{\pi}(Q^2)=-\frac{1}{16\pi^2}\left(1+\frac{\alpha_s(Q^2)}{\pi}\right)\log\frac{Q^2}{Q_0^2}-
\frac{1}{48 Q^4}\exp[-\frac{4}{5\lambda_g}(\frac{1}{Q}-
\frac{1}{Q_0})]\la \frac{\alpha_s}{\pi} F_{\mu\nu}^a F_{\mu\nu}^a\ra_{Q_0}-\nonumber\\ 
&&\frac{\pi\alpha_s(Q_0^2)}{Q^6}\frac{56}{81}\exp[-\frac{8}{5\lambda_q}(\frac{1}{Q}-
\frac{1}{Q_0})]\la\bar{q}q\ra_{Q_0}^2\ .
\eae

\subsection{$\phi$-correlator}

The $\phi$-meson current is defined as $j^{\phi}_\mu=-1/3\,\bar{s}\gamma_\mu s$. If we keep the $s$-quark mass 
finite, let $m_u=m_d=0$, assume that the $s$-quark condensate at $Q=Q_0$ does not 
deviate from the ones of $u$ or $d$ quarks, then at $Q\sim Q_0$ 
the OPE of the $\phi$-correlator becomes \cite{SVZ1}
\eab
\label{phi}
&&i\int d^4 \e^{iqx}\,\la T\{ j_\mu^{\phi}(x)j_\nu^{\phi}(0)\}\ra=(q_\mu q_\nu-g_{\mu\nu}q^2)\times\left\{\frac{2}{9}\left(
-\frac{1}{8\pi^2}\left(1+\frac{\alpha_s(Q^2)}{\pi}\right)\log\frac{Q^2}{Q_0^2}+
\right.\right.\nonumber\\ 
&&\left.\left.\frac{1}{Q^4}\left[\la m_s\bar{q}q\ra_{Q_0}+
\frac{1}{24}\la \frac{\alpha_s}{\pi} F_{\mu\nu}^a F_{\mu\nu}^a\ra_{Q_0}\right]-\right.\right.\nonumber\\ 
&&\left.\left.\frac{\pi\alpha_s(Q^2)}{Q^6}\left[\la 
(\bar{s}\gamma_\mu\gamma_5\, t^a s)^2\ra_{Q_0}+\frac{2}{9}\la \bar{s}\gamma_{\mu}t^a s\sum_{q=u,d,s}\bar{q}\gamma_\mu t^a q\ra_{Q_0}
\right]\right)\right\}\nonumber\\ \ .
\eae
Building in vacuum saturation and non-perturbative running of the operator VEV's, 
the restriction $Q\sim Q_0$ does not apply anymore, and
the scalar piece in curly brackets becomes 
\eab
\label{phivs}
&&\Pi^{\pi}(Q^2)=\frac{2}{9}\left(-\frac{1}{8\pi^2}\left(1+\frac{\alpha_s(Q^2)}{\pi}\right)\log\frac{Q^2}{Q_0^2}+\right.\nonumber\\ 
&&\left.\frac{1}{Q^4}\left[\exp[\frac{4}{5\lambda_q}(\frac{1}{Q}-
\frac{1}{Q_0})]\la m_s\bar{q}q\ra_{Q_0}+\frac{1}{24}\exp[-\frac{4}{5\lambda_g}(\frac{1}{Q}-
\frac{1}{Q_0})]\la \frac{\alpha_s}{\pi} F_{\mu\nu}^a F_{\mu\nu}^a\ra_{Q_0}\right]-\right.\nonumber\\ 
&&\left.\frac{\pi\alpha_s(Q_0^2)}{Q^6}\frac{112}{81}\exp[-\frac{8}{5\lambda_q}(\frac{1}{Q}-
\frac{1}{Q_0})]\la\bar{q}q\ra_{Q_0}^2\right)\ .
\eae

\subsection{Spectral functions}

Here, we calculate the spectral functions of the respective channels 
from the OPE by taking the imaginary part of $\Pi(Q^2)$ at analytical continued 
$Q^2=-s-i\ep$ or $Q=\sqrt{Q^2}=-i\sqrt{s}$ with real $s$ and $s>0$. As in the 
conventional sum rule approach we expect to obtain information on the lowest resonances 
(but now {\sl without} the apriori {\sl assumption} of 
quark-hadron duality). Fig.\,4 shows the results of 
the calculations for $\alpha_s(Q^2_0=(2\,\mbox{GeV})^2)=0.2$ we use $m_s=120$ MeV. 
Exact vacuum saturation in the sense of {\bf 1)} is assumed with 
the commonly introduced ``correction'' factor $k\ge 1$ set equal to unity.
\begin{figure}
\vspace{8.3cm}
\includegraphics{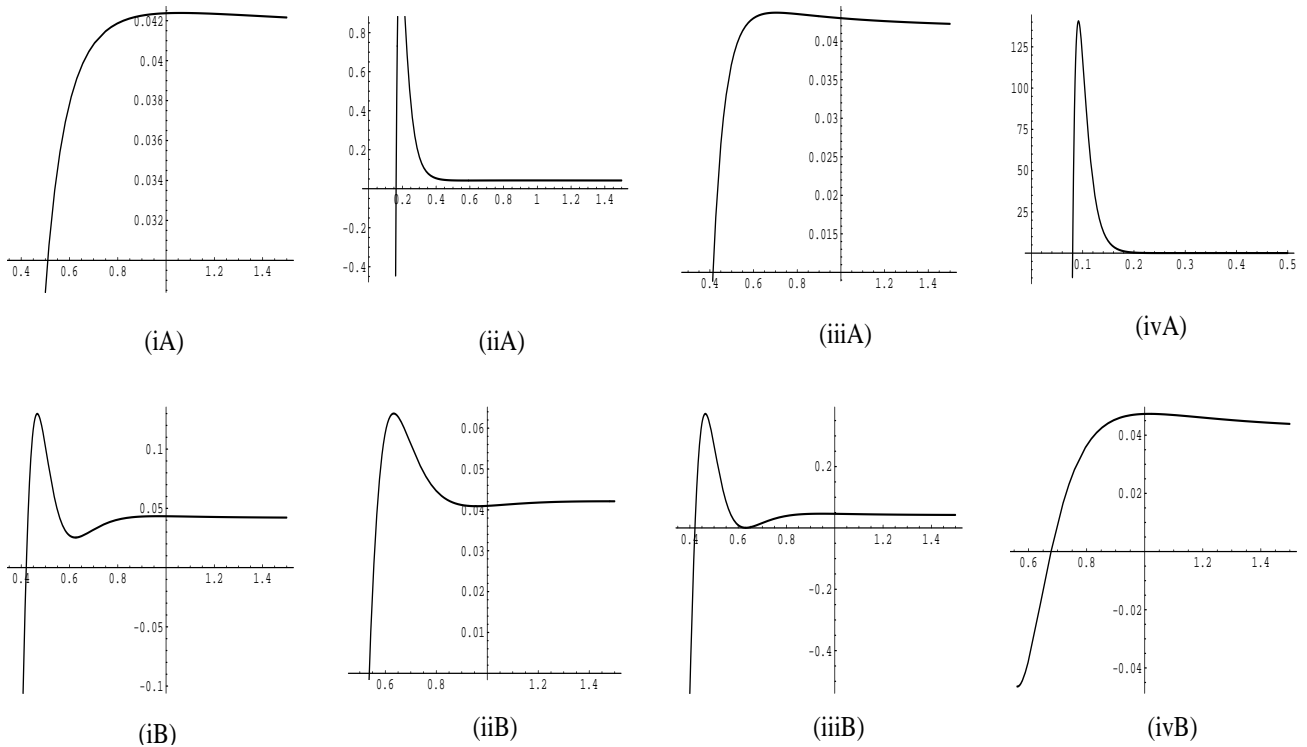}
\caption{The spectral functions $\mbox{Im}\Pi(Q^2-s-i\ep)$, $(s>0)$ 
for sets $(A)$ and $(B)$, where $(i)$, $(ii)$, $(iii)$, and $(iv)$ correspond 
to the $\rho$, $a_1$, $\pi$, and $\phi$ channels, respectively. } 
\label{} 
\end{figure}
Due to the truncation of the OPE at dimension 6 and other errors 
(large quark mass and $N_F=4$ in the lattice calculation, vacuum saturation, chiral SU(2) limit, 
and a relatively low ``fundamentality scale'' $Q_0$) we have to cut off the spectra at some 
lower bounds. Since spectral functions ought to be positive definite natural 
candidates $\sqrt{s_0}$ for these cutoffs are the first 
zeros encountered when decreasing $\sqrt{s}$.

Looking at the spectra belonging to set $(A)$ above these cutoffs, 
there is no resemblance of the measured behavior. In contrast, set $(B)$ seems 
to give a more realistic behavior: 
There is higher spectral strength at lower energy in the $\rho$ than there 
are in the $a_1$ or $\phi$ channels. In the latter we have varied the 
strange quark mass: 
For $m_s$ below 50 MeV the $\phi$ channel behaves $\rho$-like, that is, there is a 
large peak at the lower bound of the spectrum. For $m_s>$ 50 MeV the 
$\phi$-spectrum is $a_1$-like, that is, the spectral strength 
is steadily decreasing from its perturbative value down to its first zero with decreasing energy. 
Similarily, the $\pi$ channel exhibits a large concentration 
of spectral strength at low energies. 

Let us process the spectral information contained in Fig.\,4 to more quantitative statements. 
One may ask where the ``center-of-mass'' 
\eqb
\label{com}
M=\frac{\int_{\sqrt{s_0}}^{\sqrt{s_1}}(d\sqrt{s})\,\sqrt{s}\,\mbox{Im}(s)}{\int_{\sqrt{s_0}}^{\sqrt{s_1}}(d\sqrt{s})\,\mbox{Im}(s)}
\eqe
of a given spectrum within a 
low-energy domain $\sqrt{s_0}\le \sqrt{s}\le \sqrt{s_1}$ is. This is motivated by the 
experimental fact that the lowest resonance dominates the spectrum within a large range of energies. Note 
that using a narrow resonance model, the residue of this single resonance cancels in the 
definition of $M$ (ratio of moments). We may now define the universal upper 
bound $\sqrt{s_1}$ of this low-energy domain by demanding $M_\rho$ to coincide with the mass 
$m_\rho=770$ MeV of the lowest $\rho$-resonance. 
With $\sqrt{s_0^{\rho}}=420$ MeV 
eq.\,(\ref{com}) then yields $\sqrt{s_1}=1.2$ GeV. Using this, we calculate 
$M$ for the other channels with 
$\sqrt{s_0}$ always being the first zero of the respective spectral function. 
The result can be compared with the measured mass of the lowest resonance. We have
\eab
\label{M}
\sqrt{s_0^{a_1}}&=&540\ \mbox{MeV}\ ,\ \ \ \ M_{a_1}=855\ \mbox{MeV}\ ,\ \ \ \ 
\frac{M_{a_1}}{m_{a_1}}=\frac{855}{1260}\sim 0.68\ ;\nonumber\\ 
\sqrt{s_0^{\pi}}&=&420\ \mbox{MeV}\ ,\ \ \ \ M_{\pi}=680\ \mbox{MeV}\ ,\ \ \ \ 
\frac{M_{\pi}}{1/2[m_{\pi}+m_{\pi}(1300)]}=\frac{680}{720}\sim 0.94\ ;\nonumber\\ 
\sqrt{s_0^{\phi}}&=&680\ \mbox{MeV}\ ,\ \ \ \ M_{\phi}=960\ \mbox{MeV}\ ,\ \ \ \ 
\frac{M_{\phi}}{m_{\phi}}=\frac{960}{1020}\sim 0.94\ .
\eae
Due to the distinguished role of the $\pi$-meson as a Goldstone-boson 
we have taken along the next pion resonance ($\pi(1300)$) 
assuming their residues to be equal. 
The differences between the measured 
(still using the narrow resonance model) and the 
computed ratios of moments is about 30\% for the $a_1$ channel and less than 
10\% level for $\pi$ and $\phi$ channels. 
We do emphasize at this point that conventional OPE's would have given the 
same value for $M$ in all channels 
(apart from small perturbative corrections to the Wilson coefficients). 

Why do the shapes deviate considerably from the experimental ones (as a general feature, 
they seem to be shifted to lower energies)? Although the use of the small correlation length in set $(B)$ was motivated by 
ambiguities concerning vacuum saturation we 
do believe that the truncation of the OPE at 
dimension 6 is the major source 
of deviation. It is quite plausible that higher operator dimensions 
introduce shorter and shorter effective correlation lengths and 
therefore higher mass scales to govern 
the operator VEV's at low resolution \cite{PRL}. 
Viewed in this context, the use of a small 
correlation length at dimension 6 may be an 
effective way to simulate higher mass dimensions. 
Apart from this there are, of course, the unresolved problems 
linked to vacuum saturation as such and the 
way it is being implemented as they were discussed in the previous section.

\section{Implications for the calculation of $\Delta \Gamma_B$}

Motivated by the occurrence of a large (effective) 
mass scale in the previous section we discuss in this section how the notion of non-perturbative coarse 
graining of local operator averages may influence the calculation of nonleptonic, 
inclusive width differences in neutral $B$-meson systems. 

To be specific, we take the example of the $B_s$-$\bar{B}_s$ system 
as it was treated in refs.\,\cite{Nierste}. 
On the level of an effective weak Hamiltonian, which is obtained by 
integrating out the heavy bosons $Z^0, W^{\pm}$ 
of the fundamental theory (also including pQCD corrections) and which, omitting Cabibbo suppressed contributions, 
is of the form 
\eqb
\label{H}
{\cal H}_{eff}=\frac{G_F}{\sqrt{2}} V^*_{cb}V_{cs}\left(\sum_{i=1}^6 c_i O_i + c_8 O_8\right)\ ,
\eqe
the width difference $\Delta \Gamma_{B_s}$ between the mass eigenstates $\left.\left|B_{H/L}\right.\ra$ can  
be calculated  by virtue of the optical theorem as (relativistic normalization of states)
\eqb
\label{wi}
\Delta \Gamma_{B_s}=\frac{1}{2 M_{B_ s}}\,\la\bar{B}_s\left|\mbox{Im}\, i\int d^4x\, T\, {\cal H}_{eff}(x){\cal H}_{eff}\right|B_s\ra\ .
\eqe
The first six $\Delta B=1$ operators of eq.\,(\ref{H}) are of the four quark type, 
for example $O_1=(\bar{b}_{\alpha} c_{\beta})_{V-A}
(\bar{c}_{\beta} s_{\alpha})_{V-A}$, and $O_8=g/(8\pi^2)\,m_b\bar{b}_{\alpha}\sigma^{\mu\nu}(1-\gamma_5)
F^a_{\mu\nu}t^a_{\alpha\beta}s_\beta$. 

To be able to handle eq.\,(\ref{wi}) one expands in position space 
the T-product into an OPE which contains $\Delta B=2$ operators, for example 
\eqb
\label{B=2}
Q=(\bar{b}_{\alpha} s_{\alpha})_{V-A}
(\bar{b}_{\beta} s_{\beta})_{V-A}\ .
\eqe
In view of the subsequent $x$ integration the OPE only makes 
sense if the momentum $q$ of the external 
states is in the euclidean (or space-like) domain. 
So for a meaningful calculation formally an analytical continuation of the time-like momentum squared $q^2\sim m_b^2$ 
of the heavy mesons at rest to $q^2=-Q^2<0$ must be performed. After a determination 
of the Wilson coefficients in the euclidean the result is analytically continued back to $q^2\sim m_q^2$. 
Thereby, one has to deal with two renormalization scales $\mu_{1,2}$. 
The former is due to the expansion of the fundamental, electroweak (together with QCD corrections) 
interaction into the effective Hamiltonian, whereas the latter comes in via the separate 
scale dependence of the $\Delta B=2$ operators and Wilson coefficients. It is claimed in refs.\,\cite{Nierste} 
that in the $\Delta B=2$ OPE the dependence on $\mu_1$ 
is almost cancelled on the level of the Wilson coefficients at the same order in $\alpha_s$. 
Since the $\overline{\mbox{MS}}$--scheme was used in \cite{Nierste} 
the $\overline{\mbox{MS}}$ matrix elements had 
to be matched to the matrix elements obtained in lattice renormalization at the 
low lattice renormalization point 2$\sim$ GeV. Subsequently, they were run up to $\sim m_b$.  

Keeping the scale of the 
matrix element fixed in a purely perturbative 
renormalization group evolution, the transcendental dependences on the 
external momentum scale are powers 
of logarithms due to their resummation. Violating local duality, 
we have seen in the previous sections that these logarithmic 
dependences do not introduce resonance structure in 
the spectral functions of 
light-quark channel vacuum correlators. 
If the scale of the process is comparable to the inverse (effective) correlation length then 
there is a much stronger dependence of the product of Wilson coefficient and 
matrix element of a local operator than the logarithmic one. Along the lines of section 3 
it would therefore be important to measure the 4 quark correlators 
corresponding to the $\Delta B=2$ matrix elements of eq.\,({wi}) 
in order to decide whether mass scales (inverse correlation lengths) 
occur which are dangerous for the heavy quark expansion. 
This should also be done for contributions of higher 
dimensional operators formally corresponding to higher powers in $1/m_b$. 
After all, the inverse, 
effective correlation length, used in the OPE's of vacuum correlators in 
light-quark channels (set $(B)$ in section 4), is not 
too small compared to $m_b$ (10/3 GeV $\not\ll$ 4.5 GeV).

\section{Summary}

In this paper we investigated the consequences of non-perturbative 
coarse graining of operator VEV's as it was proposed in ref.\,\cite{PRL}. 
The focus was on light-quark correlators. 

After a brief review of OPE coarse graining, based 
on the knowledge of non-perturbatively calculated, 
gauge invariant $n$-point functions, we addressed the 
issue of vacuum saturation in the case of 
so-called disconnected diagrams. In particular, 
it was pointed out that there are ambiguities in the way approximations 
to the general prescription are implemented for coarse graining 
VEV's of local 4 quark operators. 
Using the simplest option for vacuum saturation, the machinery was applied to the light-quark correlators 
in the $\rho$, $a_1$, $\pi$, and $\phi$ channels. The spectral functions of the respective channels were 
calculated from the OPE by analytical continuation to time-like external momenta. 
Using lattice data on the gauge invariant field strength and scalar quark correlators, 
the spectra were found to be far off their experimentally measured behavior. 
With a 10 times smaller correlation length at dimension 6 the basic phenomenological 
features turned out to be 
contained in the spectral functions. However, in general the spectra are 
shifted to lower energies and resonance information 
is vague in the $a_1$ and $\phi$ channels. 
This may be a consequence of excluded higher mass 
dimensions and the treatment of the VEV's of 
4 quark operators. Despite the obvious shortcomings 
a calculation of the ratios of the first and zeroth moments of the spectral distributions in the 
$a_1$, $\pi$, and $\phi$ channels gave quite realistic results after fixing 
this ratio in the $\rho$ channel. Finally, we discussed the potential 
impact non-perturbative coarse graining may have on the 
theoretical determination of the inclusive, 
nonleptonic $\Delta \Gamma$ in $B_s$-meson decays.

\section*{Acknowledgements}    

The author would like to thank Uli Nierste for a stimulating conversation. Financial support 
from CERN's theory group during a research stay in June are gratefully acknowledged. 
The author is indebted to V. I. Zakharov for numerous useful discussions and valuable comments.

\bibliographystyle{prsty}

\begin{thebibliography}{10}

\bibitem{SVZ}
M. Shifman, A. Vainshtein, and V. Zakharov, Nucl. Phys. B{\bf 147}, 385 (1979).

\bibitem{Bigi}
I. Bigi and N. Uraltsev, hep-ph/0106346.

\bibitem{CNZ}
K. G. Chetyrkin, S. Narison, V. I. Zakharov, Nucl. Phys. B{\bf 550}, 353 (1999).
    
\bibitem{Shifman0}
M. Shifman, {\sl Theory of Preasymptotic Effects in Weak Inclusive Decays}, 
in Proc. Workshop on {\sl Continuous Advances in QCD}, ed. A. Smilga 
(World Scientific, Singapore, 1994)

\bibitem{Shifman1}
M. Shifman, in {\sl Particles, Strings and Cosmology}, eds. J. Bagger {\sl et al.}, 
World Scientific Singapore 1996, hep-ph/9505289.\\ 
B. Chibisov, R. D. Dikeman, M. Shifman, and 
N. Uraltsev, Int. J. Mod. Phys. {bf A 12}, 2075 (1997), hep-ph/9605465.\\  
B. Blok, M. Shifman, and D. Zhang, Phys. Rev. {\bf D}{\bf 57},  2691  (1998); (E) {\bf D}{\bf 59},  019901  (1999).

\bibitem{PRL}
R. Hofmann, hep-ph/0109007.

\bibitem{Dosch}
H. G. Dosch, Phys. Lett. B {\bf 190}, 177 (1987).\\ 
H. G. Dosch and Yu. A. Simonov, Phys. Lett. B {\bf 205}, 339 (1988).

\bibitem{DiGiacomo}
M. D'Elia, A. Di Giacomo, and E. Meggiolaro, Phys. Lett. B{\bf 408}, 315 (1997).\\ 
A. Di Giacomo, E. Meggiolaro, and H. Panagopoulos, Nucl. Phys. B{\bf 483}, 371 (1997).\\ 
M. D'Elia,  A. Di Giacomo, E. Meggiolaro, Phys. Rev. {\bf D}{\bf 59}, (1999) 054503.
     
\bibitem{Meg}
M. D'Elia,  A. Di Giacomo, E. Meggiolaro, Phys. Rev. {\bf D}{\bf 59}, (1999) 054503.

\bibitem{Weisz}
A. Bode et al., hep-lat/0105003.

\bibitem{SVZ1}
M. Shifman, A. Vainshtein, and V. Zakharov, Nucl. Phys. B{\bf 147}, 448 (1979). 

\bibitem{Nierste}
M. Beneke, G. Buchalla, C. Greub, A. Lenz, and U. Nierste, Phys. Lett. B {\bf 459},  631.\\ 
U. Nierste, hep-ph/0009203.

\end{thebibliography}

\end{document}